\documentclass[%
 aip,
 pop,
 amsmath,amssymb,
 reprint,
]{revtex4-1}
\usepackage{siunitx}
\usepackage{graphicx}% Include figure files
\usepackage{dcolumn}% Align table columns on decimal point
\usepackage{bm}% bold math
\usepackage[utf8]{inputenc}
\usepackage[T1]{fontenc}
\usepackage{mathptmx}
\usepackage{etoolbox}
\usepackage{hyperref}
\usepackage{xcolor}
\usepackage{subcaption}
\makeatletter
\def\@email#1#2{%
 \endgroup
 \patchcmd{\titleblock@produce}
  {\frontmatter@RRAPformat}
  {\frontmatter@RRAPformat{\produce@RRAP{*#1\href{mailto:#2}{#2}}}\frontmatter@RRAPformat}
  {}{}
}%
\makeatother

\begin{document}

%\preprint{LLNL-JRNL-XXXXXX}

\title{Measurements of Laser-Driven Plasma Expansion into Hohlraum-Relevant Background Gas}
% \title{Measurements of Laser-Driven Plasma Expansion into Hohlraum-Relevant Helium Gas Fills}

\author{S.~Hilsabeck}
\affiliation{University of California, Berkeley, California 94720, USA}

\author{S.~Dannhoff}
\affiliation{Massachusetts Institute of Technology, Cambridge, Massachusetts 02139, USA}

\author{C.~A.~Walsh}
\author{M.~Sherlock}
\author{G.~D.~Sutcliffe}
\affiliation{Lawrence Livermore National Laboratory, Livermore, California 94550, USA}

\author{E.~R.~Tubman*}
\affiliation{University of California, Berkeley, California 94720, USA}
\email{e.r.tubman@berkeley.edu}

\date{\today}

\begin{abstract}
Experiments at the OMEGA EP laser facility were designed and executed to study plasma expansion into hohlraum-relevant gas fills (0.3-0.6 mg/cc of helium), providing a surrogate platform for investigating hohlraum wall blow-off, non-local transport, and magnetized plasma effects. We observe well-defined density features and filamentary structures as laser-driven copper plasma expands into a low-Z background gas. Shadowgraphy resolves sharp density features over time and reveals fine-scale filamentation in the laser spot region with characteristic transverse scales of \SIrange{10}{100}{\micro\meter} near the foil surface. Proton radiography provides sensitivity to path-integrated magnetic fields and density modulations throughout the bubble volume. We extract the bubble expansion, as a function of time, for two gas pressures, 350~psi (producing 0.3~mg/cc equivalent conditions) and 700~psi (0.6~mg/cc equivalent conditions), and compare the measured propagation to magnetohydrodynamic simulations performed with Gorgon and HYDRA. While the large-scale shape of the bubble is well reproduced by both codes, the time-dependent expansion rate shows significant discrepancies (20–50$\%$ faster) compared to experimental observations between 1 and 3 ns. This leads to increasingly larger differences in bubble sizes at later times. The optical measurements of bubble expansion and evolution of small-scale structures point to additional constraints required for Biermann-battery field generation, thermal transport, and instability growth in hohlraum-relevant plasmas, to ensure accurate, predictive modeling of gas-filled hohlraums.
\end{abstract}

\maketitle

\section{Introduction}
\label{sec:intro}

In indirect-drive inertial confinement fusion (ICF), high-power laser beams deposit energy in the interior of a high-$Z$ hohlraum to generate a symmetric x-ray drive for capsule implosion. Laser irradiation of the hohlraum walls also produces high-temperature plasmas that expand inward into a low-$Z$ gas fill. The resulting laser-driven plasma bubbles influence the propagation of the incident laser, the energy deposition and ultimately the x-ray conversion efficiency \cite{Ralph2018HohlraumDynamics,Callahan2018SymmetricDriveLimits, Farmer2017}. Inner beams traverse the expanding plasma created by the outer beams, where density gradients and electromagnetic fields can perturb laser propagation and degrade x-ray drive symmetry. In gas-filled hohlraums, helium is typically used to regulate wall blow-off \cite{Hall2017}. However, the detailed dynamics of bubble expansion, including the structure of the compressed gas layer, the role of self-generated magnetic fields, and the impact of filamentary instabilities are not yet fully understood. Until now, experimental platforms have not isolated the expansion of a single plasma bubble into a hohlraum-relevant gas fill.

Strong temperature and density gradients near the laser spot generate magnetic fields via the Biermann battery mechanism\cite{Biermann1950, Kulsrud2008CosmicMagFields},
\begin{equation}
\frac{\partial \mathbf{B}}{\partial t}
= \frac{k_B}{e}\frac{\nabla T_e \times \nabla n_e}{n_e}
\label{eq:biermann}
\end{equation}
and can modify electron heat transport, steepen density gradients, and influence the overall expansion dynamics\cite{Schoeffler2018}. Extended-MHD and radiation-hydrodynamics codes such as Gorgon\cite{Chittenden2004} and HYDRA\cite{Marinak2001} typically treat transport using flux-limited diffusion and approximate models of magnetized conduction. While these approaches capture many features of large-scale flows, the experiments presented here indicate that they do not fully reproduce the temporal expansion of the plasma bubble or the fine-scale structure of instabilities in hohlraum-relevant plasmas.

A number of prior studies have used optical diagnostics and proton radiography to observe filamentary structures and magnetic fields around laser spots on foils in a vacuum\cite{Li2009,Huntington2015,Campbell2020}. These works have demonstrated that self-generated fields can reach tens of tesla and that small-scale density modulations may arise from combinations of thermal, kinetic, and Weibel-like mechanisms\cite{Li2009,Huntington2015,Sutcliffe2022}. However, in many cases, the connection between such small-scale structures and the macroscopic expansion of hohlraum wall plasmas is not well understood.

While electromagnetic fields generated within hohlraums have been investigated experimentally using proton radiography in both gas-filled and vacuum hohlraums\cite{Li2009,Li2012,Pearcy2025}, the geometric complexity makes it difficult to isolate individual expansion features and determine their underlying physical mechanisms. We have designed a new platform that allows us to more simply probe and understand laser-driven plasma expansions into hohlraum-relevant gas fills, and benchmark simulation codes, such as HYDRA and Gorgon. Irradiation of a foil launches an expanding copper plasma, normal to the target surface, into a helium gas producing a compressed helium layer bounded by the Cu–He interface and an outward-propagating shock front. This interaction forms distinct regions of ablated copper plasma, shocked and compressed helium, and unshocked ambient helium, throughout which instabilities and fine-scale structures are observed to develop.

Of particular interest to this work are the bubble expansion rate, the structure and instabilities that evolve within the compressed helium layer, and the appearance of fine-scale filamentation near the foil surface. These features can influence energy transport, magnetic-field generation, and the overall morphology of the expanding plasma, yet are either under-resolved or absent in magnetohydrodynamic simulations. 

In this paper, we present the experimental study of plasma bubble expansion in helium-filled hohlraums, designed to mimic the conditions of hohlraum wall blow-off at gas densities of 0.3 and 0.6~mg/cc. We irradiate copper foils at the OMEGA EP facility to produce expanding plasma bubbles that are diagnosed with shadowgraphy, interferometry, and dual-axis proton radiography. The platform extends previous datasets to hohlraum-relevant gas densities and we will discuss a side-by-side comparison of experimental observables with results from Gorgon and HYDRA simulations. In addition, we observe that late-time interferometry is affected by perturbations that expand into the reference leg of the Wollaston interferometer\cite{Howard2018} which complicates direct phase retrieval in this regime. This motivates the development of forward-modeling approaches for future comparisons between simulated and experimental interferograms.

\section{Experimental platform}
\label{sec:platform}

\begin{figure}
  \centering
  \includegraphics[width=1\columnwidth]{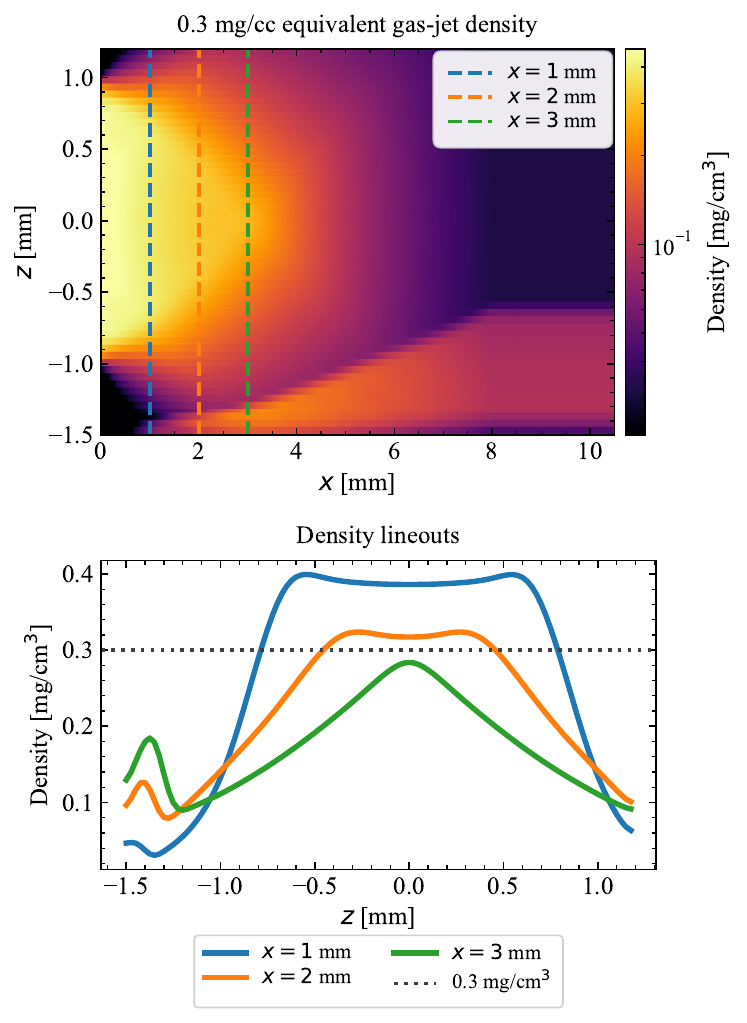}
  \caption{Gas-jet density profile corresponding to a 350 psi He backing pressure to produce the 0.3 mg/cc equivalent conditions. The density distribution is approximately symmetric about the jet centerline, with features consistent with a shock rebounding from the foil surface. Vertical lineouts are taken at $x=1, 2$, and 3 mm away from the exit of the gas jet nozzle, where $x=2$ mm corresponds to the nominal laser focal position. The resulting profiles illustrate the evolution of the gas density as it propagates across the foil region.}
  \label{fig:gasjet}
\end{figure}

Experiments were carried out on the OMEGA-EP laser system at the Laboratory for Laser Energetics (LLE). The primary targets were Cu foils of thickness \SIrange{2}{5}{\micro\meter}, supported by \SI{25}{\micro\meter} aluminum backing foils and positioned at the exit of a bent gas jet nozzle. The foil thickness was varied to optimize probing capabilities, with thinner foils producing sharper proton radiography features, while thicker foils were used at later times to ensure that the expanding plasma remained dominated by ablated Cu. The gas jet produced a background helium gas with a density profile as shown in Fig. \ref{fig:gasjet}  \cite{McMillen2024}. At the location of the laser focus, 2 mm along the gas nozzle axis, the He densities of $\sim 0.3$ and $\sim 0.6$~mg/cc were formed from nominal backing pressures of 350 and 700~psi, respectively. Within a focal spot (\(\sim\)\SI{750}{\micro\meter} diameter), the gas density is approximately uniform. As the plasma expands to diameters exceeding 1 mm, it samples regions of the gas jet with modest density variations. Simulations incorporating the gas jet profile indicate that these variations have a negligible effect on the plasma expansion over the timescales considered. Over the duration of the experiment, the ejected gas moves only a few microns (v$\approx 1.5 $ km/s) \cite{Hansen2018}, and can therefore be treated as stationary over the timescales of interest.

\begin{figure}
  \centering
  \includegraphics[width=1\columnwidth]{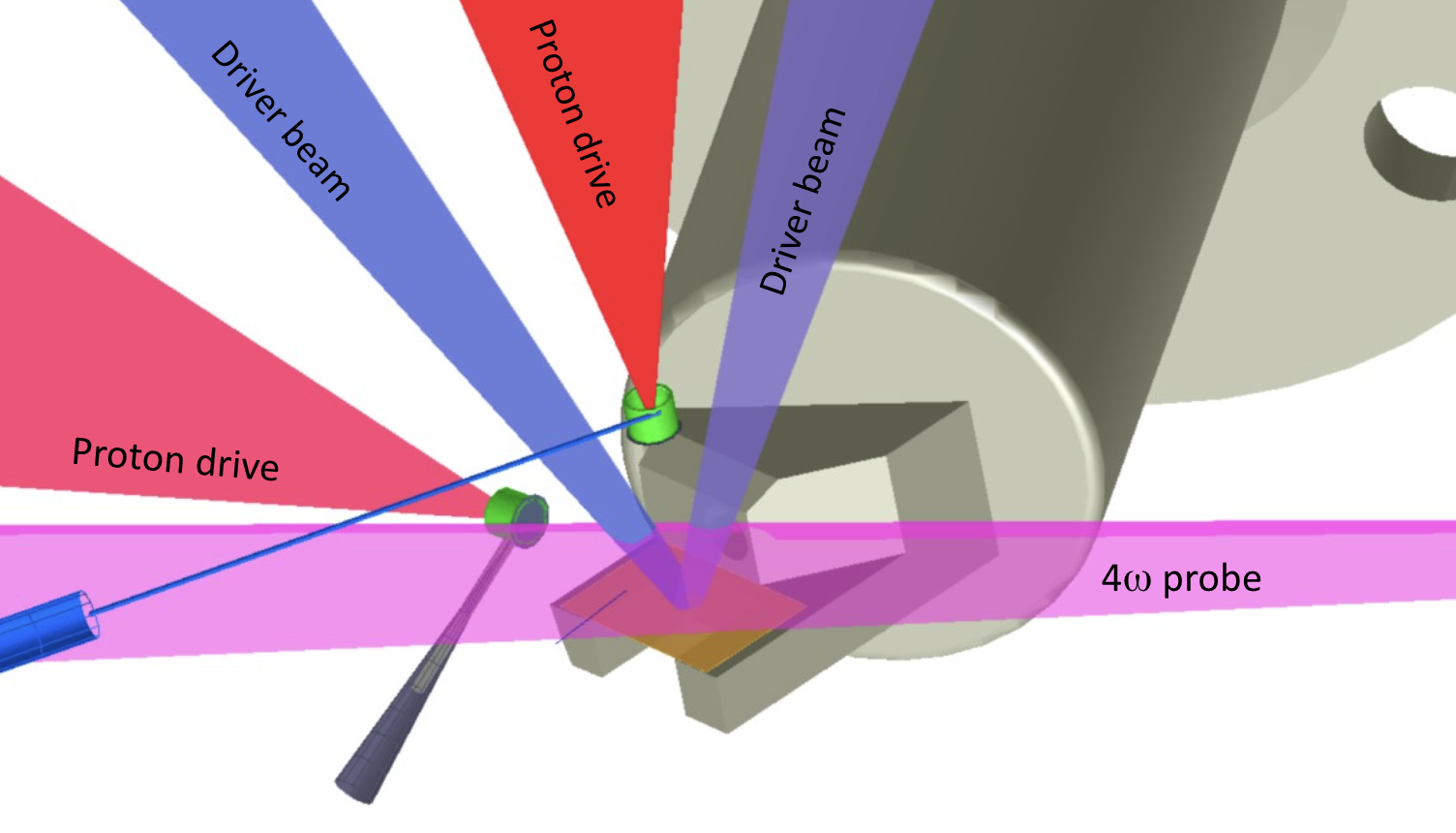}
  \caption{Diagram of the experimental setup. 3$\omega$ drive beams (purple) are incident on a thin Cu foil, launching a plasma bubble into a He background gas. Two proton backlighter targets provide side-on and face-on proton radiographs, while a 4$\omega$ probe (pink) passes through the plasma parallel to the foil surface for shadowgraphy and interferometry measurements.}
  \label{fig:VisrRad}
\end{figure}

Frequency-tripled drive beams at $\lambda = 351~\text{nm}$ irradiated the foil. Beams delivered approximately 4200--4300~J in a 4~ns pulse. Two beams were overlapped to form spots of $\sim 750~\mu$m diameter, corresponding to peak intensities of $I \sim 2.5\times 10^{14}~\text{W/cm}^2$. The timing of a $4\omega$ probe beam relative to the drive beams was varied between 1--4.0~ns, capturing both the initial acceleration of the plasma expansion and the onset of deceleration as the bubble formed a compressed He gas layer. The orientation of the foil and gas jet was chosen such that the probe beam propagated parallel to the foil surface and perpendicular to the bubble expansion, enabling measurements sensitive to the expanding plasma structures. Additionally, two short pulse lasers (3 ps, 200 J each) were focused onto \SI{10}{\micro\meter} thick Cu backlighter foils, located inside proton backlighter tubes capped with \SI{3}{\micro\meter} tantalum foil to provide shielding to the proton foil from debris, x-rays and gas. The backlighter foils were located 7 mm from the main foil target and produced protons of up to 30 MeV via target normal sheath acceleration (TNSA)\cite{Mackinnon2004, Wilks2001} to image the electromagnetic fields both side-on and face-on relative to the foil surface. The protons were captured onto radiochromic film (RCF) detector stacks located 78 mm from the main target. The resulting spatial modulation of the proton fluence on the detector provides a map of the path-integrated electromagnetic field structure\cite{Schaeffer2023ProtonImaging}.

\section{Shadowgraphy and Wollaston interferometry}
\label{sec:diagnostics}
A frequency-quadrupled 263~nm (4$\omega$) probe beam was used to perform shadowgraphy and interferometry\cite{Froula2012}. The shadowgraphy system is sensitive to the second derivative of the refractive index\cite{Settles2001}, providing high-contrast visualization of the steep density gradients and shock boundaries within the plasma. This makes shadowgraphy particularly effective for tracking the compressed gas regions, the expanding plasma bubble, and for resolving fine-scale features in the plasma density.

\begin{figure}
  \centering
  \includegraphics[width=0.8\columnwidth]{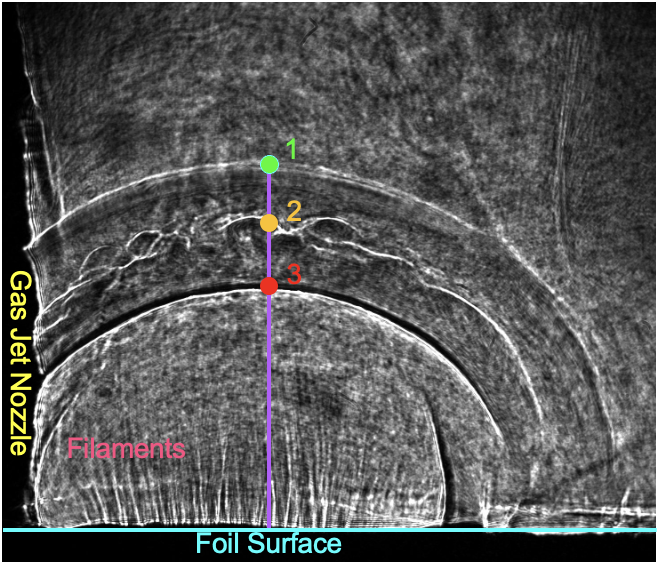}
  \caption{Shadowgraphy image of the expanding plasma bubble in the 0.6 mg/cc He background gas case at 3 ns. Three distinct expansion features, labeled 1, 2, and 3, are identified along the axial direction. The image also shows filamentary structures within the expanding plasma, the He gas jet, and at the foil surface. The purple line indicates the axial lineout location used to extract the expansion positions.}
  \label{fig:bubble_features}
\end{figure}

The probe beam, after passing through the plasma and image-relay optics, is split onto various cameras to record interferograms, shadowgraphs and angular filter refractometer images\cite{Haberberger2013}. The system can resolve $\sim\SI{1}{\micro\meter}$ structures in the plasma (corresponding to $\sim\SI{7}{\micro\meter}$ in the diagnostic plane) across the entire field of view. Images were recorded on a 27~mm~$\times$~27~mm, 16-bit CCD with \SI{13.5}{\micro\meter} pixels, yielding a $\sim 2.5$~mm field of view.

\begin{figure}
  \centering
  \includegraphics[width=0.7\columnwidth]{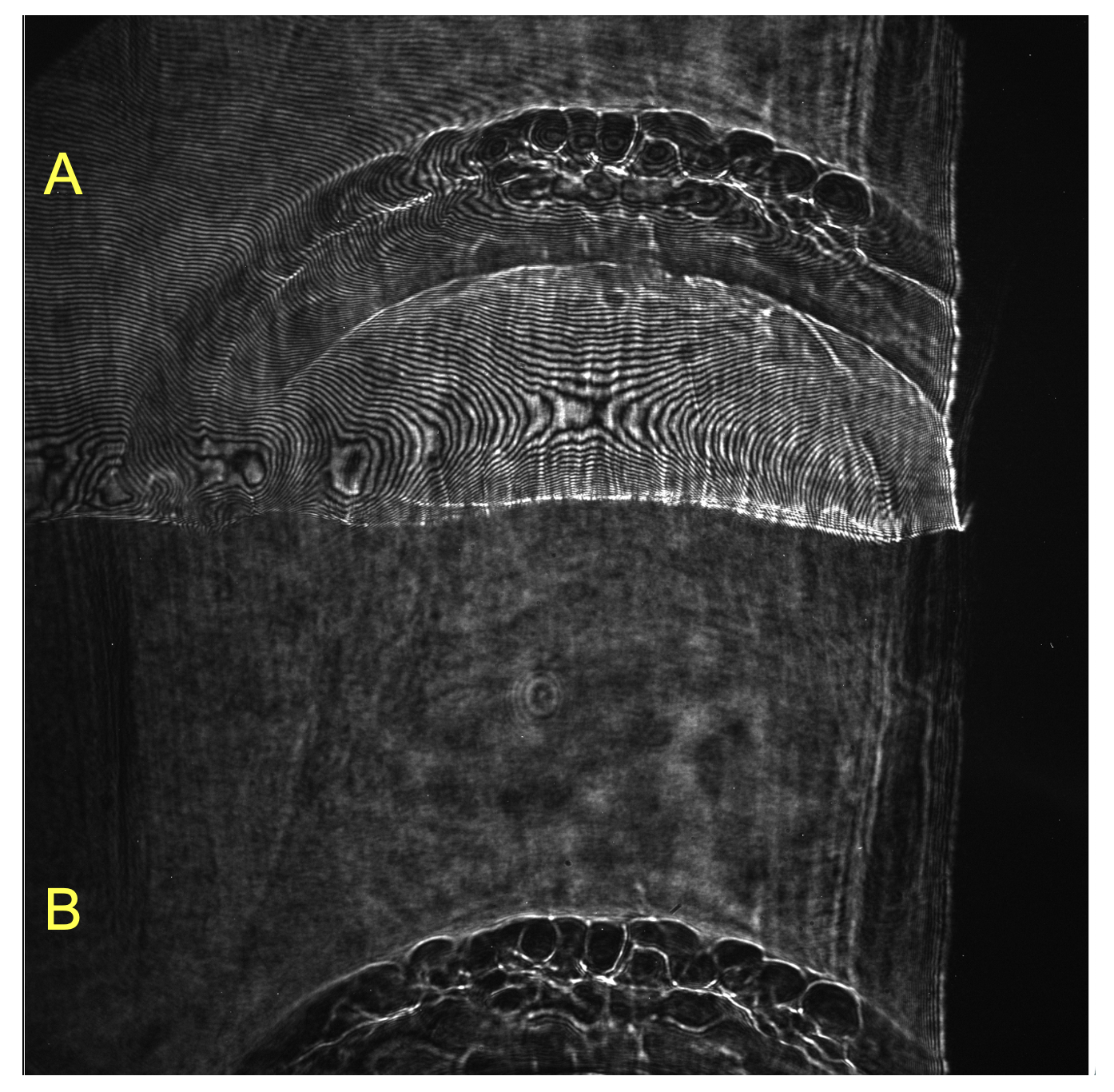}
  \caption{Interferogram at 2~ns for the 0.6 mg/cc He case.
  Fringes are perturbed as the plasma occupies both Leg~A (signal path) and Leg~B (reference path), indicating that the expanding plasma has filled the entire interferometer aperture and compromised the reference-leg stability.}
  \label{fig:int}
\end{figure}

The Wollaston interferometer is a self-referencing interferometer which splits the 4$\omega$ probe into a signal and reference leg. Even at early times ($\sim 1$~ns), the expanding Cu plasma and background gas start to refract the probe beam in both the signal and the reference legs of the interferometer. Perturbations to the reference arm (Fig.~\ref{fig:int}) prevent accurate phase retrieval\cite{Howard2018}. In this regime, fringe curvature cannot be directly inverted into an absolute phase map, and interferometry is therefore limited to qualitative interpretation and relative density measurements.

Quantitative measurements of the shock and interface position reported in this work were obtained from shadowgraphy. A forward model of the Wollaston interferometer (Sec. \ref{sec:future}) is under development to compare simulated density fields with observed fringe distortions, including reference-arm perturbations at times when both interferometer arms contain plasma.

\section{Tracking expansion features from shadowgraphy}

\begin{figure}[!t]
  \centering
  \includegraphics[width=0.85\columnwidth]{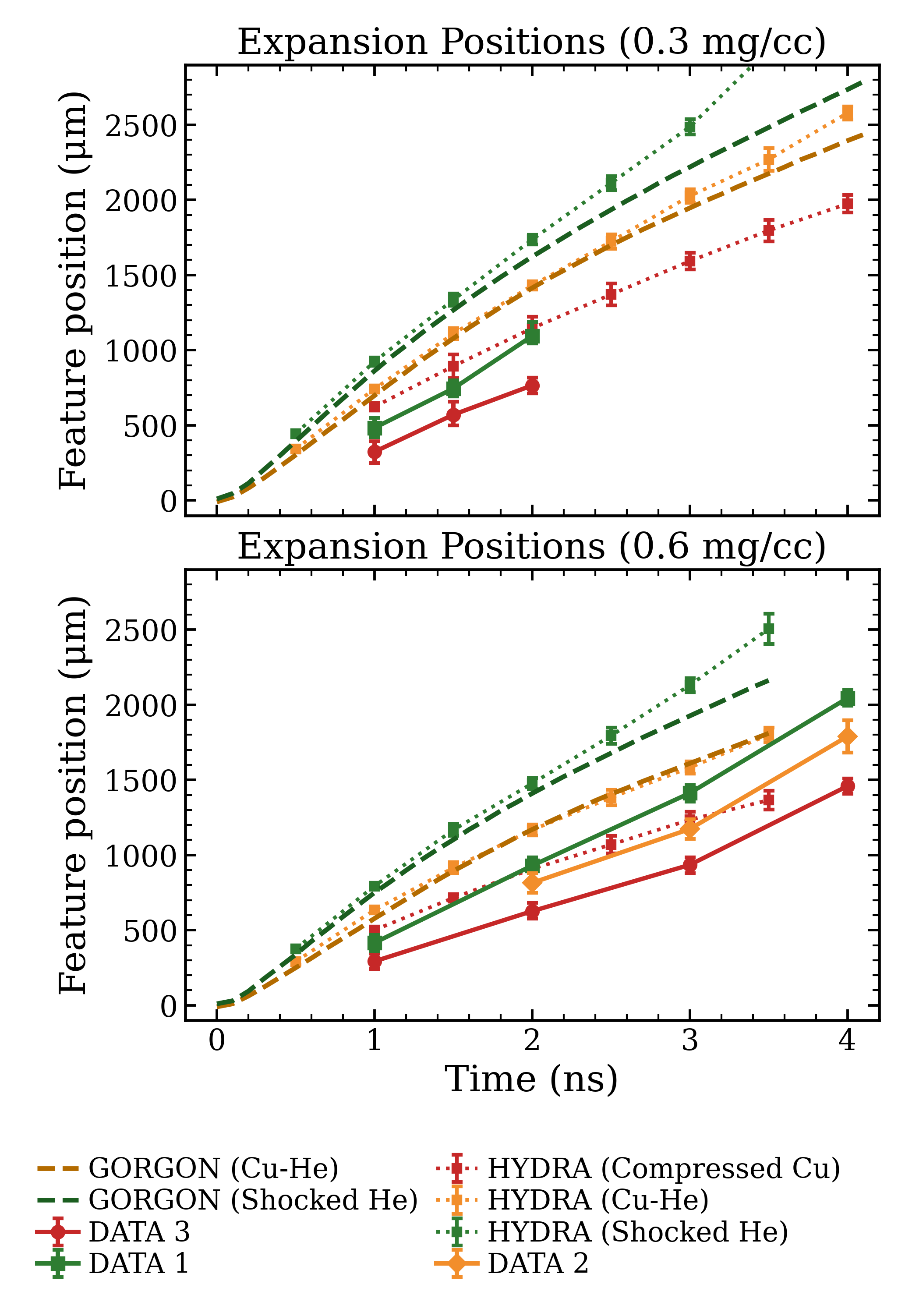}
  \caption{Expansion positions as a function of time for the 0.3 mg/cc (top) and 0.6 mg/cc (bottom) He background gas cases. Experimental measurements are shown for the tracked features labeled 1 (green), 2 (orange; 0.6 mg/cc only), and 3 (red). These are compared with Gorgon simulations (dashed lines), which predict the Cu–He interface (orange) and the shocked He front (green), and HYDRA simulations (dotted lines with square markers), which additionally predict the compressed Cu front (red). At 0.3 mg/cc, only the outermost (1) and innermost (3) experimental features are resolved, whereas all three features are resolved at 0.6 mg/cc. Overall, both simulation codes predict faster expansion than observed experimentally, although the level of agreement depends on the interface and background gas density.}
  \label{fig:positions}
\end{figure}

\begin{figure}[t]
  \centering
  \includegraphics[width=0.85\columnwidth]{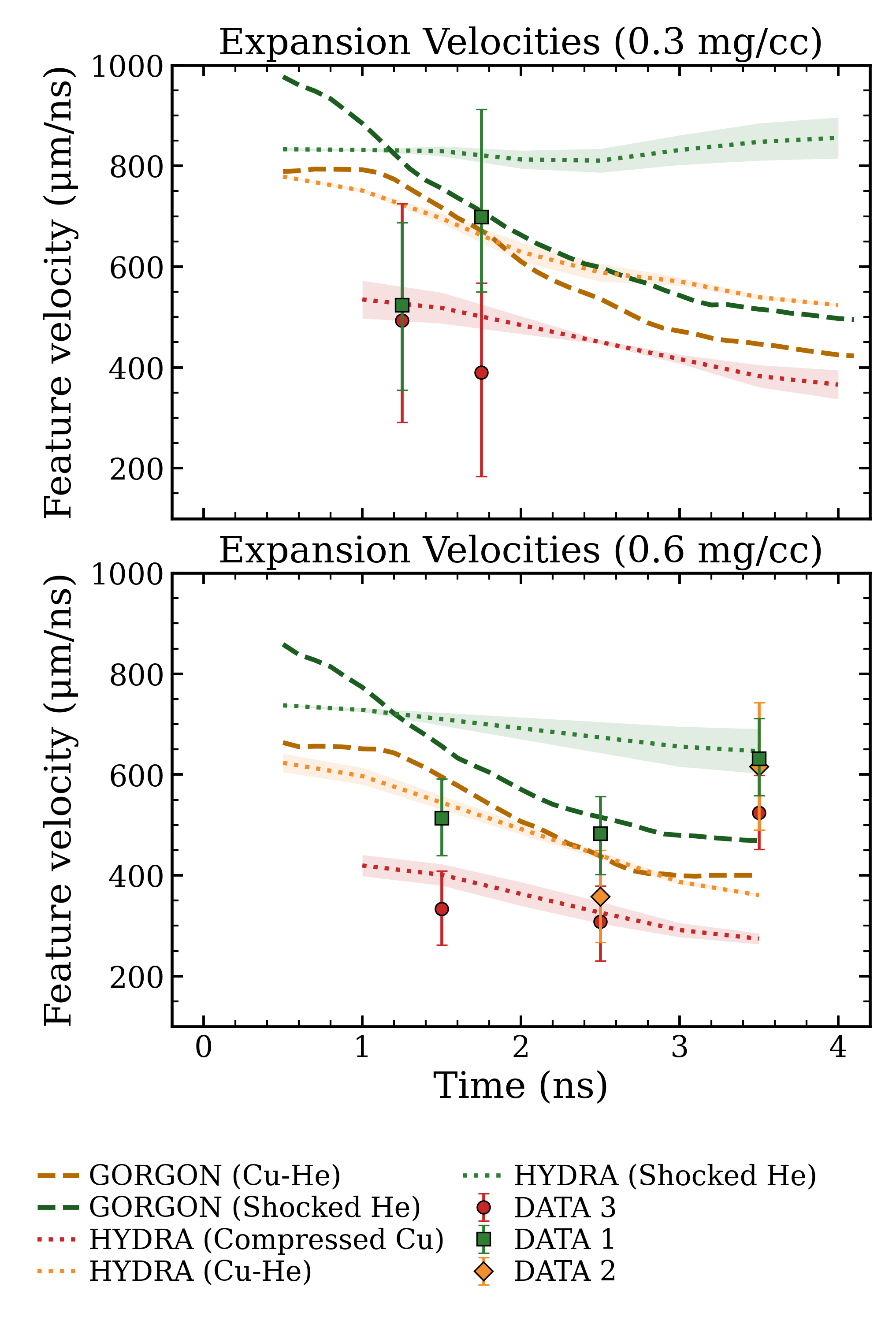}
  \caption{Expansion velocities, calculated from the measured and simulated feature positions, for the 0.3 mg/cc (top) and 0.6 mg/cc (bottom) He background gas cases. Experimental measurements are shown for the tracked features labeled 1 (green), 2 (orange; 0.6 mg/cc only), and 3 (red). These are compared with Gorgon predictions for the Cu–He interface (orange triangles) and shocked He front (green triangles), and HYDRA predictions for the compressed Cu front (red squares), Cu–He interface (orange squares), and shocked He front (green squares). The simulations reproduce the Cu–He interface velocities reasonably well but generally predict higher velocities for the shocked He front than observed experimentally, particularly at later times.}
  \label{fig:velocities}
\end{figure}

Expansion feature positions were extracted from intensity lineouts taken along a single direction normal to the foil surface as shown in Fig. \ref{fig:bubble_features}. Along each lineout, distinct intensity gradients correspond to boundaries of steep density gradients. For the 0.3 mg/cc case, two expansion features were consistently resolved, while for the 0.6 mg/cc case, three distinct features could be identified. The distance of each feature from the initial foil surface defined its expansion position at each probing time. Because the plasma becomes over-critical close to the foil, the original foil surface location is less well defined in the experimental images. Instead, its position was determined from reference images acquired before plasma formation. Local turbulent and filamentary structures produce a finite spread in the apparent location of each boundary. The innermost and outermost plausible boundary locations were therefore used to define the uncertainty bounds on each position measurement. Reference shots taken before the foil was ablated indicated that shot-to-shot variations in the initial foil position, arising from foil non-flatness and target alignment, had deviations of up to $\SI{50}{\micro\meter}$. This variation is included in the uncertainty in Fig. \ref{fig:positions} and Fig. \ref{fig:velocities}.

\begin{figure*}
\centering
\begin{subfigure}{\textwidth}
    \centering
    \includegraphics[width=0.8\textwidth]{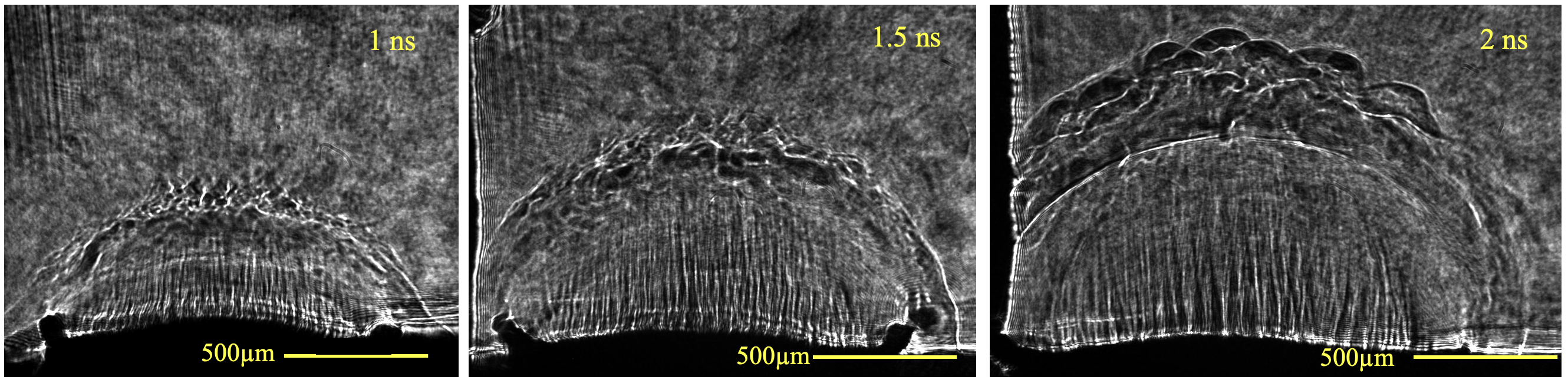}
    \caption{}
    \label{fig:350psi_expansion}
\end{subfigure}

\vspace{0.5em}

\begin{subfigure}{\textwidth}
    \centering
    \includegraphics[width=0.8\textwidth]{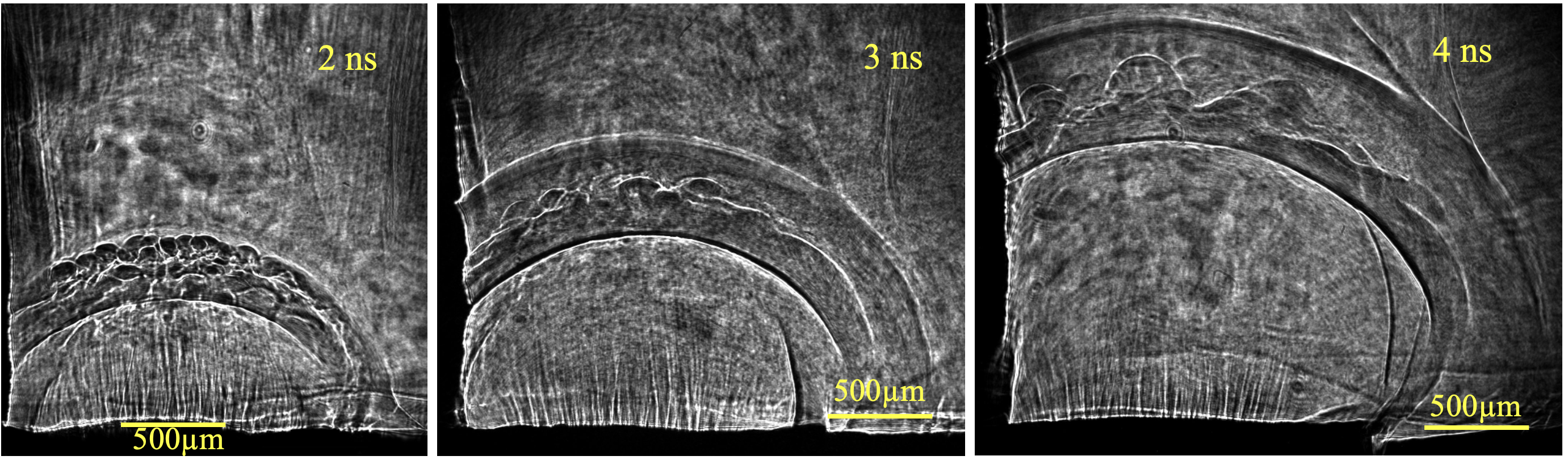}
    \caption{}
    \label{fig:700psi_expansion}
\end{subfigure}

\caption{Shadowgraphy of Cu expanding into (a) 0.3 mg/cc He at 1, 1.5, and 2 ns and (b) 0.6 mg/cc He at 2, 3, and 4 ns. The expanding plasma produces multiple distinct density features that become increasingly well resolved with time. By 3 ns, three expansion features are clearly visible in the 0.6 mg/cc case, accompanied by fine filamentary structures near the foil surface.} \label{fig:shadowgraph}
\end{figure*}

At early times (1 ns), when intensity gradients are not sharply defined because of turbulence near the foil, as seen in Fig. \ref{fig:350psi_expansion}, the feature positions were determined by first averaging the intensity over a $\SI{20}{\micro\meter}$ window parallel to the foil surface.

\section{Experimental results}
\label{sec:results}

\subsection{Qualitative morphology}

Under drive irradiation, the Cu foil ablates and launches an expanding plasma bubble into the He gas. The interaction produces several distinct density features that are visible in the shadowgraphy images shown in Fig.\ref{fig:bubble_features}. In the 0.6 mg/cc case, three distinct expansion features were resolved, while in the 0.3 mg/cc case only the innermost and outermost features are clearly visible because of probing earlier in time. The outer feature, labeled 1, is formed by a shock-like perturbation being launched into the background He gas. Feature 3 likely corresponds to the compressed copper front. The images also show turbulent structures at the location labeled 2 between the innermost and intermediate features, which may be indicative of hydrodynamic instability, potentially consistent with Rayleigh–Taylor-like growth developing at the Cu-He interface. Without measurements that are able to spatially resolve the presence of He and Cu at this time it is challenging to identify clear material regions. Near the foil surface, fine-scale filamentary structures appear within the expanding plasma, with transverse scale lengths in the range of \SIrange{10}{100}{\micro\meter}. The filament spacing does not appear to vary with time or background gas density, and in both the 0.3 mg/cc and 0.6 mg/cc cases the filamentation remains confined near the foil. The confinement of these filaments to the laser-irradiated region suggests they may arise from laser-driven processes, such as thermal or ponderomotive filamentation.

\begin{figure}
  \centering
	\includegraphics[height=0.25\textheight]{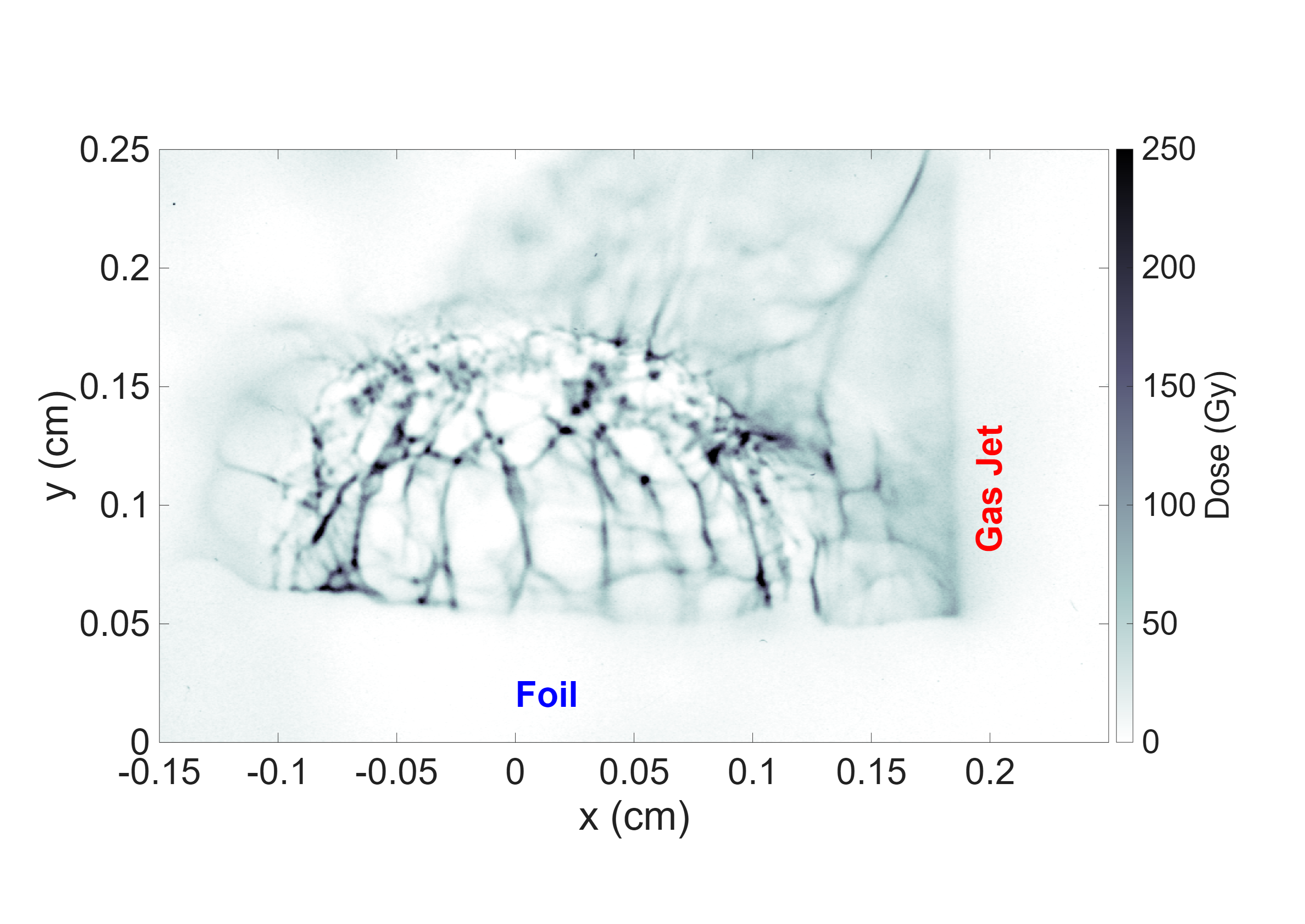}
%	\subcaptionbox{\label{fig:deflangle}}{\includegraphics[height=0.2\textheight]{proton_angledefl.png}}
  \caption{Proton radiography of a calibrated image of the raw proton dose recorded on the RCF using 18.6 MeV protons. The protons probe side-on, across the foil surface at 2 ns, with a 0.3 mg/cc He background gas. This radiograph indicates electromagnetic structures throughout the bubble volume. Shadows of both the foil and the gas jet can be seen in the image.}\label{fig:rawprad}
\end{figure}

\subsection{Electromagnetic field deflections}
Proton radiographs exhibit modulations in proton fluence throughout the bubble volume, as seen in Fig. \ref{fig:rawprad}. Narrow bands of enhanced fluence and broader regions of reduced fluence result from proton deflections by electromagnetic field structures within the plasma. The turbulent structures observed in shadowgraphy, associated with strong density gradients, may seed magnetic fields through mechanisms such as the Biermann battery mechanism, Eq.~(\ref{eq:biermann}), the Weibel or magnetothermal instabilities. 

The radiograph shown in  Fig. \ref{fig:rawprad} shows the electromagnetic structures that have evolved at 2 ns, when the copper plasma expands into 0.3 mg/cc He background gas. The apparent size of the bubble is $\sim1200 ~\mu$m; however, this does not account for shadowing by the extended foil, which may block protons deflected downwards, effectively clipping the image. There might also be lensing effects from strong electric fields near the target surface distorting and enhancing structures close to the foil. Both caustic and void structures are also apparent in the radiographs, requiring careful analysis to distinguish their origins. Preliminary simulations indicate that the smaller filamentary and web-like features are likely associated with magnetic-field deflections, while larger electric fields are primarily generated near regions of sharp discontinuity. These results will be explored in future studies.

Electromagnetic fields provide evidence that self-generated magnetic fields and kinetic anisotropies may play an important role in the dynamics of the expanding plasma. Further analysis using particle-tracking and synthetic-radiography tools such as PROBLEM\cite{Bott2017} will enable a more detailed interpretation of the proton deflections and the path-integrated electromagnetic fields. Comparing the characteristic spatial scales observed in proton radiography and shadowgraphy will also help determine whether the two diagnostics are probing the same underlying instabilities and provide insight into their respective seed mechanisms.

\subsection{Bubble expansion at 0.3~mg/cc and 0.6~mg/cc}

Using the shadowgraphy images, the expansion features identified in Sec.~IV~A were tracked. For the 0.3~mg/cc (350~psi) He gas case, shown in Fig.~\ref{fig:350psi_expansion}, two distinct expansion features were consistently resolved throughout the experiment. The outermost feature moved from 0.483~mm to 1.094~mm above the target foil between 1 and 2~ns, corresponding to an average expansion velocity of 0.611~mm/ns. Over the same interval, the innermost feature expanded from 0.324~mm to 0.765~mm, giving an average velocity of 0.442~mm/ns. Finite-difference velocities derived from the measured positions indicate a rapid initial expansion followed by a decrease in growth rate, although additional probe times would be required to confirm this trend. A quadratic fit to the position--time data yields a negative acceleration, consistent with the expansion slowing as the plasma propagates into the surrounding He.

At the higher-density 0.6~mg/cc (700~psi) He background gas case, shown in Fig.~\ref{fig:700psi_expansion}, three distinct expansion features were resolved. The outermost feature propagated from 0.416~mm to 0.930~mm between 1 and 2~ns, corresponding to an average velocity of 0.514~mm/ns, approximately 15\% lower than in the 0.3 mg/cc case, consistent with the increased inertia of the denser background gas. Beyond 2~ns, the velocities of the outermost, intermediate, and innermost features remain approximately constant within the measurement uncertainties (10--20\%). This behavior indicates that, over the 1--4~ns interval, the expansion remains predominantly drive-dominated while the main laser beams are on, with only modest deceleration due to the swept-up background gas.

Plasma bubbles reach larger expansion distances and higher inferred velocities over the 1--2~ns interval in the 0.3 mg/cc background He, consistent with the expectation that the higher-density 0.6 mg/cc background gas provides greater confinement and therefore reduces the expansion rate.

\subsection{Comparison between experiment and simulation}

The multi-group radiation-hydrodynamics codes HYDRA\cite{Marinak2001} and Gorgon\cite{Chittenden2004} were used to model this experiment. 

The HYDRA simulations were performed in a quasi-2D cartesian slab and used the profile in Fig.\ref{fig:gasjet} to model the background gas (scaled for the 0.6 mg/cc experiments) with a 0.1 mg/cc density floor. Beam pointing is aligned with the plane of the slab, with incident angles at the foil matching those in the experiment. The simulations shown here use a thermal flux limiter of 0.15, which is typical of a `high flux limit' for hohlraum calculations\cite{farmer_spheres}, though varying the electron thermal flux limiter had little effect on the simulated feature locations. Although HYDRA has magnetohydrodynamic (MHD) capabilities\cite{Koning2011}, the inclusion of Biermann magnetic fields and Nernst advection similarly had little impact on the simulated feature locations; the HYDRA simulations shown here omit MHD calculations. 

The Gorgon results complement those of HYDRA by utilizing a 2D cylindrical geometry, which more accurately replicates the off-normal rarefaction of the bubble. However, Gorgon assumes the gas profile is uniform and uses a normal incident laser beam. Gorgon uses a new flux limiter formulation that is more consistent with kinetic calculations\cite{walsh_FL} and limits heat-flow and Nernst transport simultaneously. The Gorgon simulations shown here include Biermann magnetic flux generation and Nernst transport down temperature gradients\cite{ciardi,walsh_prl}, and use updated magnetic transport coefficients\cite{sadler, Walsh_2021}. In addition, low-resolution three-dimensional GORGON simulations incorporating the measured gas jet density profile showed no significant change in the interface velocities.

\begin{figure}
  \centering
  \includegraphics[width=1\columnwidth]{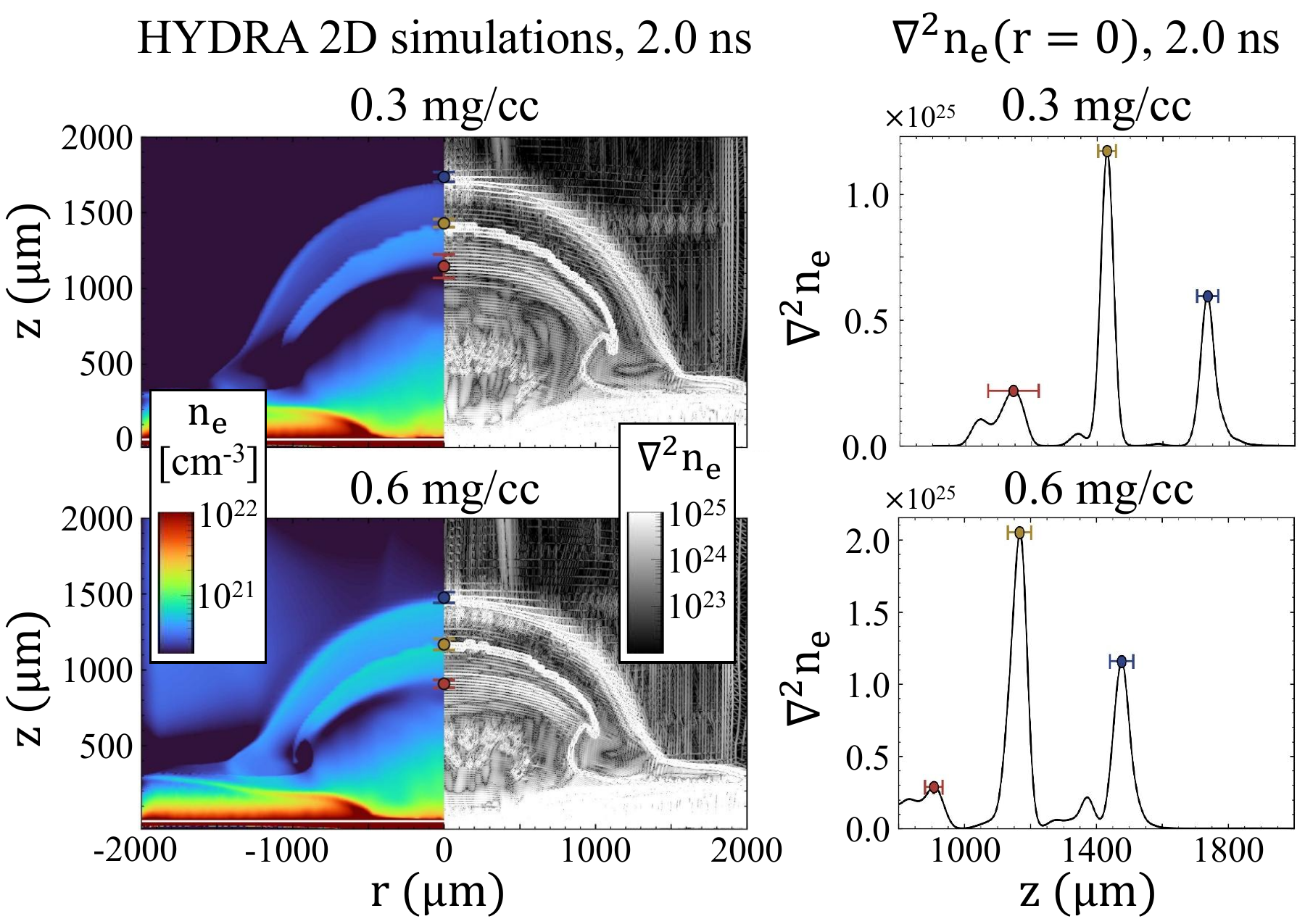}
  \caption{HYDRA electron density at 2 ns for 0.3 mg/cc (top) and 0.6 mg/cc (bottom) modeled gas profile, shown with the Laplacian of the density, $\nabla^2 n_e$. The density plots in the first column (left of $r=0$) show the expanding plasma bubble and layered structures, including the Cu plasma and the compressed He layer. The Laplacian plots in the first column (right of $r=0$) enhance sharp density gradients, producing contrast analogous to shadowgraphy. Three peaks are isolated in $\nabla^2 n_e$ lineouts taken along z at $r=0$ (right column), complicating the identification of the corresponding density features in the shadowgraphy.}
  \label{fig:hydra}
\end{figure}

The simulations are first used to identify the prominent density features. Fig.\ref{fig:hydra} shows the HYDRA-simulated electron density at 2 ns using the  0.3 mg/cc He background gas profile. The Laplacian of the electron density is also shown, which indicates which regions are most sensitive to the shadowgraphy diagnostic. An axial lineout is included to further emphasize the peaks. Starting closest to the foil, the three simulated density features are identified as follows:

\begin{itemize}
    \item Compressed Cu front (red). The inner boundary edge of the compressed Cu region, where the ablated Cu is piling up behind the Cu-He interface.
    \item Cu--He interface (yellow). The material boundary separating the Cu plasma from the compressed He layer.
    \item Shocked He front (blue). The outward-propagating shock into the ambient He background gas.
\end{itemize}

Most apparent is the absence of a clear third boundary in the 0.3 mg/cc experimental data and Gorgon simulations, which is attributed to a weaker copper compression front. In the HYDRA simulations this gradient is launched by a compressed Cu layer propagating back into the incoming ablation plasma. Because the experimental measurements do not provide material discrimination, the identity of the observed boundaries must be inferred from their evolution and comparison with the simulations. The outermost boundary, particularly evident at 4 ns in the 0.6 mg/cc case, is attributed to the shock propagating into the background helium gas. At early times, this boundary marks the outer extent of the turbulent region, as can also be seen qualitatively in the interferometry shown in Fig. \ref{fig:int}. The innermost boundary observed in the shadowgraphy is attributed to the Cu–He interface marking the edge of the expanding copper plasma. Turbulent features first develop at this interface before growing with time, while the compressed helium layer subsequently launches an additional shock into the background gas.

Both HYDRA and Gorgon broadly reproduce the qualitative evolution of the bubble but systematically over-predict the expansion rate. At early times the disagreement is modest, but it grows with time. For the 0.3 mg/cc case, the outer shocked He front is measured at $Z = 0.930~\mathrm{mm}$ at $2~\mathrm{ns}$, whereas Gorgon predicts $Z = 1.170~\mathrm{mm}$ ($+26\%$) and HYDRA predicts $Z = 1.356~\mathrm{mm}$ ($+46\%$) in the hydrodynamic configuration and $Z = 1.438~\mathrm{mm}$ ($+55\%$) in the Biermann configuration. The corresponding average expansion velocities over the 1--2~ns interval are $0.514~\mathrm{mm/ns}$ in the data, compared to $0.592~\mathrm{mm/ns}$ for Gorgon, $0.608~\mathrm{mm/ns}$ for HYDRA hydrodynamic, and $0.663~\mathrm{mm/ns}$ for HYDRA Biermann. At later times using a 0.6 mg/cc gas background, Gorgon continues to over-predict the expansion (e.g., $Z = 1.942~\mathrm{mm}$ at $3~\mathrm{ns}$ compared to $1.412~\mathrm{mm}$ in the data, $+37\%$). Differences between experiment and simulation are largest near 2--3~ns, where the codes consistently produce feature velocities above those observed experimentally. In addition, neither model generates filamentation at the observed \SIrange{10}{100}{\micro\meter} scales in the Cu plasma or compressed He layer, nor do they reproduce the complex pattern of proton fluence modulations inferred from radiography.

\section{Discussion}
\label{sec:discussion}

The experimental observations presented here highlight several key discrepancies in current radiation-hydrodynamics and extended-MHD modeling of hohlraum-relevant plasma expansion. The simulations consistently produce bubbles that are larger than those measured experimentally and do not reproduce the turbulent structures observed in the compressed He region or reproduce the filamentary structures observed near the target surface, indicating investigations with simulations employing kinetic approaches are required. 

Laser drive energy was investigated as a possible source of the discrepancy between the simulated and experimental plasma expansion. Applying a 70\% drive multiplier in the HYDRA simulations produced smaller plasma bubbles, but similar expansion rates to the nominal simulations, and these remained in disagreement with the experimental results. Gorgon simulations also show similar trends. This behavior suggests that heat flow from the Cu plasma into the He is too strong in the models. The impact of self-generated magnetic fields produced by the Biermann term could also inhibit cross-field conduction and maintain steeper temperature gradients, reducing the energy available to drive the expansion. If this suppression is underrepresented, the simulations will evolve more rapidly than the data. The simulated expansion exceeding the experimental measurements suggests that physics, such as magnetized or nonlocal thermal transport, may be missing and is likely needed to more accurately capture the evolution of the compressed He layer and its influence on bubble propagation.

The shadowgraphy and proton radiography also show fine-scale filamentation in the Cu plasma close to the target surface, whereas neither Gorgon nor HYDRA generates these structures at these scales. The turbulent structures located in the compressed He region are also absent within simulations. The smaller filaments and turbulent structures may originate from a combination of thermal and kinetic mechanisms, including Weibel-like modes driven by strong heat-flux anisotropies as well as laser-filamentation processes. Because standard radiation-hydrodynamics codes do not include kinetic anisotropy or collisionless physics, the absence of these features is expected. However, their presence in the experiment implies modifications to local transport and magnetic-field generation that are not captured in a flux-limited diffusion model, and these effects may influence the global expansion as well\cite{Huntington2015}.

\section*{Kinetic ion diffusion estimates}

Kinetic effects, notably ion diffusion and finite-mean-free-path effects, may be playing a role in determining the location of the Cu and He interfaces. Estimates of the relevant mean free path and diffusion length scale can be obtained from the plasma variables provided by the GORGON simulations in the vicinity of the Cu–He interface to assess whether Cu propagation into the He may be responsible for the turbulent structures. The ion-ion slowing mean-free-path $\lambda_{s}$\cite{ref_NRL} can be found by assuming Cu ions of charge state $Z^{*}=28$ collide with the He ions ($Z^{*}=2$, $n_{e}=2.56\times10^{26}~\mathrm{m}^{-3}$) with kinetic energies comparable to the typical ion temperature ($\sim$1.2 keV), resulting in $\lambda_s \approx \SI{0.3}{\micro\meter}$. The simulations reveal a thin layer of very hot ions generated by PdV work and although the number of ions in this layer is small, their mean-free-path can be estimated by repeating the calculation but assuming a kinetic energy of 12 keV, resulting in $\lambda_s \approx \SI{1.1}{\micro\meter}$. The perpendicular scattering collision lengths ($\lambda_{\perp}$) are similarly small. Both of these estimates are well below the values needed to explain the observed long scale-length features in the compressed gas region of the experiments, indicating that effects such as interpenetration are unlikely to explain the observed results. In addition, the diffusion length scale of Cu ions into the He plasma can be estimated from similar plasma variables obtained from the simulations. Assuming that the two species are initially separated by a sharp contact discontinuity, the concentration $c\left(x,t\right)$ of Cu ions at a time $t$ later is given by

\begin{equation}
c\left(x,t\right)=\frac{1}{2}\left\{ 1-\mathrm{Erf}\left(\frac{x}{2\sqrt{Dt}}\right)\right\} 
\end{equation}

where the diffusion coefficient $D$ is well established (see, e.g. \cite{ref_SM}) and is given in terms of the scattering mean-free-path. The diffusion length scale, $\Delta x_{D}$, is defined as the distance between the locations where $c\left(x,t\right)=0.9$ and $c\left(x,t\right)=0.1$. After approximately 1 ns of diffusion, $\Delta x_D \approx \SI{7.2}{\micro\meter}$. After approximately 3 ns of diffusion, $\Delta x_D \approx \SI{12.5}{\micro\meter}$. Again, these estimates appear too small to account for the observed experimental discrepancies.

\section{Conclusions}
\label{sec:conclusions}
We have presented a new experimental study of plasma expansion into hohlraum-relevant He gas fills using a Cu foil and gas jet platform at OMEGA EP. Shadowgraphy was used to measure the expansion of the plasma bubble, resolve distinct expansion features, and observe the onset of turbulent and fine-scale filamentation. For He background gases of 0.3 mg/cc and 0.6~mg/cc, the expansion positions were measured from 1--4~ns, and the inferred velocities and accelerations revealed significant differences in the expansion dynamics between the two background gas densities.

Comparison with GORGON and HYDRA simulations shows that the experimentally observed expansion features are broadly consistent with the large-scale plasma structure, including the shocked He front and the Cu–He interface. Both simulation codes reproduce the overall morphology and evolution of the expanding plasma, although they generally predict faster expansion than observed experimentally, with the level of agreement depending on the tracked feature and background gas density. HYDRA additionally predicts a compressed Cu layer at the Cu–He boundary that is not clearly resolved as a distinct feature in the shadowgraphy, whereas GORGON primarily reproduces the two experimentally observed boundaries. This discrepancy in bubble density could be important in a hohlraum environment where the inner beam propagation across the outer bubble affects radiation drive symmetry. The shadowgraphy and interferometry measurements are instead most consistent with tracking the outer He shock, the main turbulent Cu–He mixing region, and the outer Cu layer, rather than a distinct piled-up Cu front. Neither code reproduces the \SIrange{10}{100}{\micro\meter}-scale filamentation and additional turbulence evident in the shadowgraphy and proton radiography measurements. These discrepancies point toward missing physics in the present simulations, including the effects of self-generated magnetic fields, magnetized thermal conduction, nonlocal transport, and kinetic instabilities. The dataset presented here therefore provides a valuable benchmark for validating extended-MHD and kinetic models of laser-driven plasma expansion in gas-filled hohlraum environments.

\section{Future work}
\label{sec:future}

Identifying the dominant filamentation mechanism is essential because each mechanism alters heat transport, bubble expansion, and bubble morphology in different ways and is treated differently in radiation-hydrodynamic codes. If the observed filaments are Weibel-like, then electron-scale magnetic fields and anisotropic heat flux must be included to reproduce the experimental dynamics, which may explain why HYDRA and Gorgon over-predict the expansion rate. The filamentation mechanism directly informs which physics models must be incorporated into hohlraum simulations. Turbulent structures in the compressed outer layers and additional shock structures will also require further investigation to determine seed mechanisms. 

Future work will focus on completing quantitative reconstructions of path-integrated electric and magnetic fields from proton radiography, extending the Wollaston forward-model fitting to recover density profiles in partially over-dense regions, and performing systematic sensitivity studies in Gorgon and HYDRA to assess the impact of Biermann suppression, Nernst advection, and nonlocal heat flow on bubble expansion. Together, these efforts aim to improve the predictive capability of hohlraum simulations and to inform the design of future indirect-drive ICF experiments.

\begin{acknowledgments}
This work was performed under the auspices of the U.S.\ Department of Energy by Lawrence Livermore National Laboratory under Contract DE-AC52-07NA27344. The authors gratefully acknowledge the staff of the Laboratory for Laser Energetics for their support in conducting the experiments. We would also like to thank Peter Heuer and Pete Charles for running and providing the simulations of our gas jet profiles. 
\end{acknowledgments}

\section*{Data Availability}
The data that support the findings of this study are available from the corresponding author upon reasonable request.
\section*{References}
\bibliographystyle{aipnum4-1} % AIP numerical style for REVTeX 4.1
\bibliography{references}      % your .bib file

\end{document}